\documentclass[twocolumn,showpacs,preprintnumbers,amsmath,amssymb,prx]{revtex4}

\usepackage{graphicx}
\usepackage{dcolumn}
\usepackage{amssymb}
\usepackage{bm}
\usepackage{hyperref, xcolor}

\begin{document}


\title{Phase transition creates the geometry of the continuum from discrete space}

\author{Robert S. Farr$^{*\dag}$ and Thomas M. A. Fink$^{\dag}$}
\affiliation{
$^{*}$Jacobs Douwe Egberts, Banbury, OX162QU, UK\\ 
$^\dag$London Institute for Mathematical Sciences, 35a South St, 
Mayfair, London, UK
}
 \email{robert.farr@JDEcoffee.com}
 \email{tmafink@gmail.com}

\begin{abstract}\noindent
Models of discrete space and space-time that exhibit continuum-like 
behavior at large lengths could have profound implications for physics. 
They may help tame the infinities arising from quantizing gravity, and
remove the need for the machinery of the real numbers; a construct with
no direct observational support. 
However, despite many attempts to build discrete space, 
researchers have failed to produce even the simplest geometries. 
Here we investigate graphs as the most elementary discrete models 
of two-dimensional space.
We show that if space is discrete, it must be disordered, by proving that 
all planar lattice graphs exhibit a taxicab metric similar to square grids. 
We then give an explicit recipe for growing disordered discrete space by 
sampling a Boltzmann distribution of graphs at low temperature. 
Finally, we propose three conditions which any discrete model of Euclid's
plane must meet: have a Hausdorff dimension of two, support unique 
straight lines and obey Pythagoras' theorem. Our model satisfies 
all three, making it the first discrete model in which continuum-like 
behavior emerges at large lengths.
\end{abstract}


\date{Received 27th March 2019. Published 16th August 2019}

\maketitle


\section{Introduction}


The small-scale structure of space has puzzled scientists and 
philosophers throughout history. Zeno of Elea \cite{Hagar} claimed
that geometry itself is impossible because there is no consistent form
this small-scale structure can take. He argued that a 
line segment, which can be halved repeatedly, cannot ultimately be composed 
of pieces of non-zero length, else it would be infinitely long.
However, it also cannot be composed of pieces of zero length, for no matter
how many are added together, the resulting line will  
never be longer than zero. 

It is a lasting tribute to the optimism of researchers that work on geometry
nevertheless carried on.
It was not until the 19$^{\rm th}$ century -- nearly
two and a half millennia later -- that Cantor finally
resolved the paradox by defining the continuum. 
He showed that
the line must be composed not just of an
infinite number of points, but of an uncountably infinite number, 
so that the second half of Zeno's argument (a proof by induction) fails.
This uncountable infinity is described by the mathematical machinery 
of the real numbers.
The continuum is the basis for all descriptions of space and space-time,
and therefore all of theoretical physics.

In the 20$^{\rm th}$ century, Weyl \cite{Weyl} further claimed
that the continuum is the
only possible model of space. He constructed
a tiling argument, purporting to show that if space is discrete, 
Pythagoras' theorem -- or, equivalently, the Euclidean metric -- is false.
Weyl's proof, however, contains an unstated assumption which turns out
to be the key to its resolution.

\begin{figure}
\begin{center}
\includegraphics[width=0.6\columnwidth]{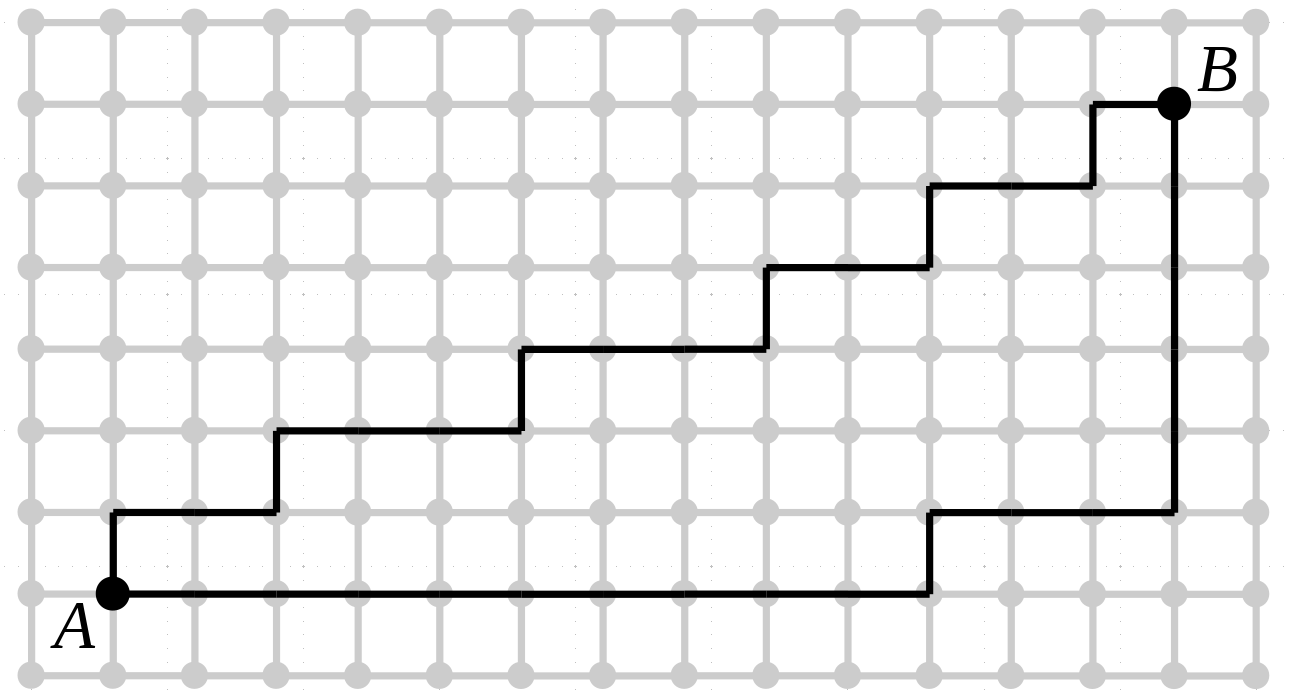}
\end{center}
\caption{\label{taxi}
\textbf{The geometry of the square grid graph.}
Two nodes $A$ and $B$ on the square grid graph are separated by 19 edges.
There are many possible shortest paths (geodesics) of length 19 edges
between the nodes, of which two are shown in black. The resemblance to 
the possible routes followed by yellow cabs in New York city inspired
the term `taxicab metric' for the measure of distance on this graph 
\cite{Taxicab}.
}
\end{figure}

Despite this long belief in the necessity of the continuum,
researchers are actively pursuing discrete \cite{Bombelli,Benincasa,Hamma}, 
or at least piece-wise flat \cite{Regge,Regge2,Ambjorn,Ambjorn2,Horava}, 
models of space and space-time, as they offer
the possibility to remove non-renormalizable infinities which arise
in simple versions of quantum gravity. 
All these models can be thought of as
graphs, where just the graph itself matters, not its embedding into another
space. The only natural \cite{Leymarie} metric in this case is graph
geodesic distance: the distance between two nodes is 
the smallest number of edges joining them.

In two dimensions, toy models of `quantum graphity' \cite{Konopka}
aim to define Hamiltonians over all graphs with a fixed number of nodes, from 
which an approximation to a smooth manifold might emerge at low temperature.
Recent attempts have aimed to produce 
planar graphs made up of triangles,
but, so far \cite{Conrady,Chen}, the low-temperature phases contain
defects, conical singularities and multiply-connected topologies, so
are unlike simple, smooth manifolds. A basic feature of such graphs is their
Hausdorff dimension, which in this context is the power with which the
number of nodes in a ball of radius $r$ grows with $r$. 
If one deliberately restricts the
ensemble of graphs under consideration to triangulations of the plane,
a further problem with graph models of two-dimensional space
is encountered:
Completely random triangulations (i.e. typical graphs chosen at random from
this ensemble, and termed `Brownian maps') 
do not even have Hausdorff dimension two. They are so crumpled
that the number of nodes in a disc of radius $r$ scales as $r^4$,
not $r^2$ \cite{leGall}.

In light of these difficulties, the prospects for building a consistent 
discrete model of even the Euclidean plane seem poor. In this Article, 
we show that it is in fact possible to discretize space.
We do three things. First, we prove that any discrete model of two-dimensional 
space must be disordered, by showing that all planar lattice graphs 
have a taxicab-like metric \cite{Taxicab}. Order is the hidden assumption in
Weyl's proof of the impossibility of discrete space.
Second, we describe a local, statistical process, with an associated 
temperature, which provides an 
explicit recipe for growing disordered graphs.
Third, we propose three tests which any model of Euclidean space must pass.
We find that graphs grown by our thermal process, at low temperature,
achieve the required properties: they have a Hausdorff dimension 
of 2, support the existence of unique straight lines, and satisfy 
Pythagoras' theorem.

\section{Lattice graphs are taxicab graphs}
\begin{figure}
\includegraphics[width=1.0\columnwidth]{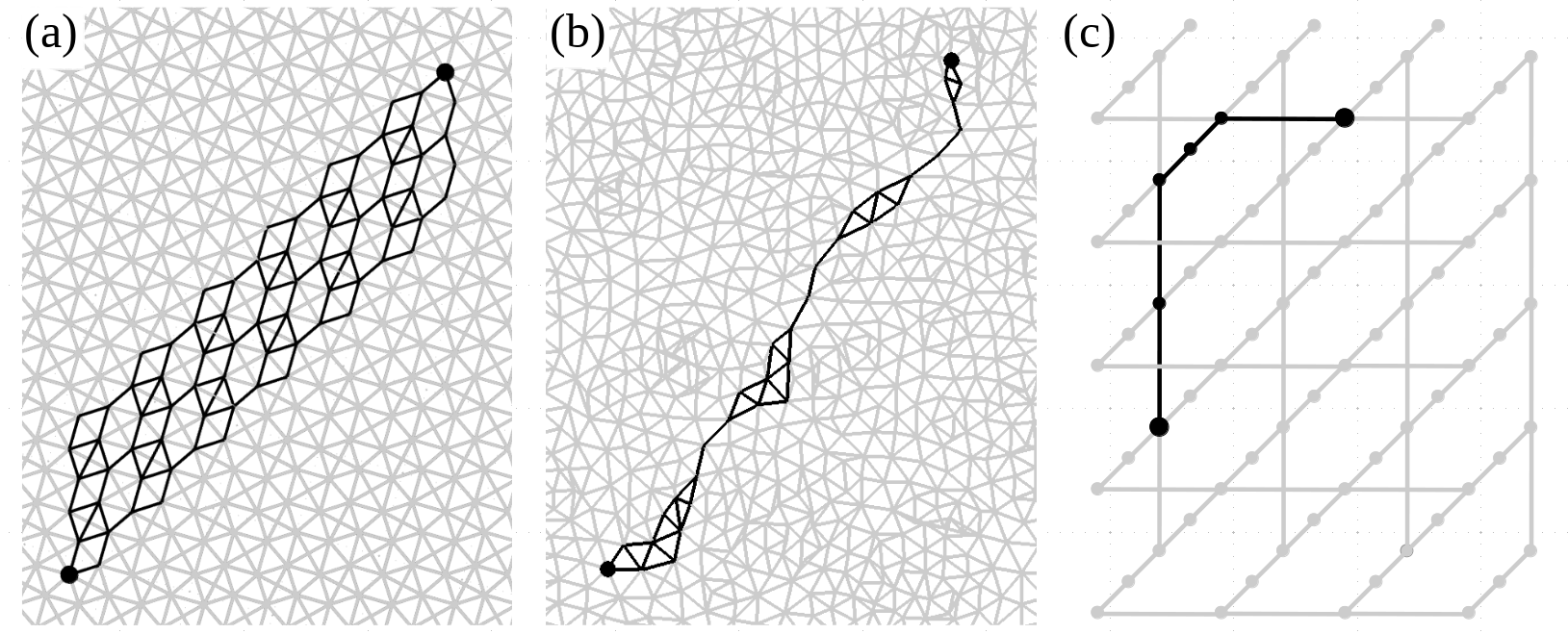}
\caption{\label{geodesic_pics}
\textbf{Geodesic confinement is not found in planar lattice graphs but is in 
planar disordered graphs.}
(a) In a doubly-periodic triangulation 
(a modified snub square tiling),
two nodes marked as circles are 22 edges apart.
We call the set of all geodesics between them (shown in black) the geodesic 
bundle, containing a number of nodes proportional to the square of the 
geodesic length.
(b) In a random triangulation, the geodesic bundle between two nodes 
22 edges apart is confined to a narrow region.
We call this phenomenon geodesic confinement. (c) A nonplanar doubly
periodic graph (all nodes shown as circles) has neither a taxicab nor 
Euclidean metric.
}
\end{figure}

The natural way to measure the distance between two nodes
on a graph is to count the edges in the shortest path which separates them.
A shortest path of this kind is called a geodesic.
It is well known that with this measure of distance,
the square grid graph has a taxicab geometry \cite{Taxicab},
where the distance between two nodes is the sum of the magnitude of the 
differences of their Cartesian coordinates (Figure \ref{taxi}). 
On this graph there are typically many geodesics between two nodes a 
distance $\lambda$ apart, each resembling an irregular staircase.
Together these form a geodesic bundle 
comprising $N_{\rm geo}\propto \lambda^{2}$ nodes.
More complex lattice graphs show a similar phenomenon 
(Figure \ref{geodesic_pics}a).

In general, any doubly-periodic planar graph must belong to one of the 
wallpaper groups, familiar from crystallography, and
be composed of unit cells containing one or more nodes.  
We prove that all doubly-periodic planar graphs have the taxicab metric,
regardless of the complexity of the unit cell.
That is to say, geodesics in all but a finite number of directions 
form broad, parallelogram bundles (the number of exceptional directions
may, however, be large for sufficiently complex unit cells \cite{Fritz}).
Such graphs therefore do not satisfy Euclid's axiom of a unique straight 
line between two points, nor Pythagoras' theorem.
Our proof is in two parts, which we call geodesic composition and 
geodesic rearrangement. We sketch the proof here, and give full
details in the Methods section.

\subsection*{Sketch of the proof}
If we have a geodesic on a graph between two nodes, 
it is clear that cutting it in two
yields two paths, each of which is a geodesic between its respective 
end nodes (were
that not so, it would be possible to create a shorter path between the
original two nodes). Even in classical geometry, however,
putting two geodesics (straight lines) end-to-end does not always give a
geodesic: they need to be parallel. The situation with graphs is more
interesting still.

Equivalent nodes
in different unit cells are said to be of the same type. We first construct
a geodesic between two nodes of
the same type, which are separated by a vector distance of
$(m,n)$ unit cells. If we choose the node type so that this is
the shortest of all such geodesics (or one of the shortest, if the
choice is not unique), then we are able to prove that many copies of
this path can be concatenated end-to-end, and the result is still a geodesic
between the now widely-separated end points.
We call this the geodesic composition property. It is not trivial, 
since it relies on the assumption of planarity; a non-planar counterexample 
is shown in Figure \ref{geodesic_pics}c.

Next, we show that a long concatenation of this single type of
geodesic can, apart from
short tails at the ends, be broken down into many alternating copies of two
different geodesics. The proof uses Dedekind's pigeonhole principle
\cite{Pigeon}, applied to the number of nodes in the unit cell. 
If $m$ and $n$ are relatively prime, these two
geodesics are not parallel. They therefore perform the role of the
coordinate directions in the square grid graph and, in the same way, can be
re-arranged in any order to produce many irregular staircase-like geodesics,
all of the same length. The set of these geodesics forms the broad geodesic 
bundle, with an area proportional to the square of 
its length -- a complete contrast to
the narrow lines required by Euclidean geometry.

\section{Growing disordered graphs}
\begin{figure}
\includegraphics[width=1.0\columnwidth]{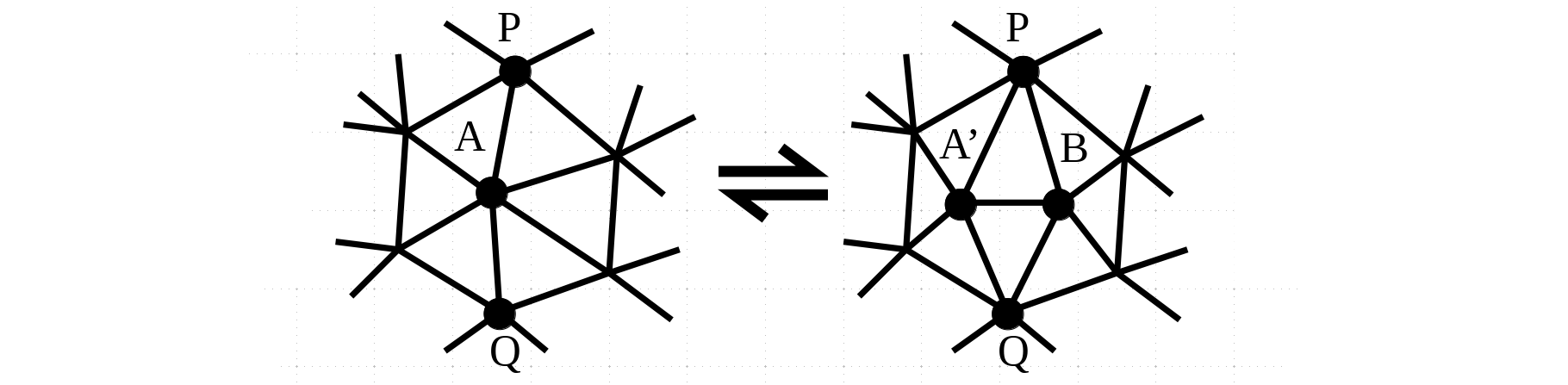}
\caption{\label{moves}
\textbf{Steinitz moves on a portion of a triangulation.}
The push move  (left to right) consists of choosing a node $A$ and two 
opposite neighbors $P$ and $Q$
(or nearly opposite, if the degree of $A$ is odd). 
Node $A$ is divided into nodes $A'$ and $B$. 
The pop move (right to left) consists of
choosing a node $A'$, and then one of its neighbors $B$, which are then
merged to a single node $A$.
In contrast to Ref.\ \cite{Steinitz}, which keeps track of triangular faces,
we avoid tetrahedra and bottlenecks smaller than 4 edges, so faces can
be assigned unambiguously, if desired. To ensure no tetrahedra are formed 
during a pop move, we make an additional check before the merger of
$A'$ and $B$: we require that
no neighbor of $A'$ that is not $P$, 
$Q$ or $B$ is connected to a neighbor of $B$ that is not $P$, $Q$ or $A'$.
}
\end{figure}

In light of the impossibility of generating Euclidean geometry
from planar lattice graphs, we turn to disordered graphs which 
triangulate the 2-sphere.
Triangulations here are graphs composed of triangles which, when 
embedded in the 2-sphere, are planar \cite{Lutz}. 
We also require that they contain no tetrahedra, so that we only need
to keep track of nodes and edges, not faces.
As a seed graph, we start from the octahedron (Figure \ref{hot_and_cold}),
a simple triangulation of the 2-sphere.
All triangulations of the 2-sphere are known to be transformable
into one another by Steinitz moves \cite{Steinitz}, illustrated in
Figure \ref{moves}, which are local, and
add (`push') or remove (`pop') nodes while preserving the property
of being a triangulation.

We grow the seed graph to a size of $N$ nodes through push moves,
and then apply $8N$ alternating push and pop moves to ensure equilibration.
This equilibration stage is necessary,
since some graph properties (such as mean node eccentricity 
-- see section \ref{testing}) change slightly,
converging (presumably to thermal equilibrium values)
after around $4N$ alternating push-pop moves moves.

Let $Z_i$ be the degree of node $i$, and $\langle Z \rangle$ the mean
degree of all nodes in the triangulation.
Because every triangular face has three edges, and every edge belongs to 
two triangles, Euler's polyhedron theorem \cite{Euler} implies that 
\begin{equation}
\langle Z\rangle = 6-12/N.
\label{meanZ}
\end{equation}

Since the integrated Gaussian curvature over a smooth, closed surface
is $4\pi$ \cite{Bonnet}, we see that 
$\kappa_i\equiv 6-Z_i$ is a natural
measure of the local, discrete equivalent 
of Gaussian curvature for the triangulation, up to a constant factor.
If we consider a patch of the graph 
consisting of $N_{\rm pat}$ nodes,
with $e$ exiting edges, 
and with a simple closed-path perimeter of length $p \ge 3$ edges, 
then we find the Euler characteristic implies the average
discrete curvature over all nodes in the patch is
\begin{equation}
\langle \kappa \rangle_{\rm pat} = (6+2p-e)/N_{\rm pat}.
\end{equation}
This can be shown by considering a new triangulation, formed from
two copies of the patch, and identifying nodes and edges on the perimeters,
then correcting for the exiting edges.
Thus a Steinitz push move locally decreases 
$|\langle\kappa\rangle_{\rm pat}|$, and
a pop move increases it. 

To create an ensemble of graphs, we first define an energy $E$ for every
graph. We then repeatedly select a random node as a candidate for a push or 
pop move, and calculate the
energy change $\Delta E$ that would result.  
We perform the move with a probability given by the Metropolis algorithm 
\cite{Metropolis} with an associated temperature $T$. Thus, the move
is always accepted if $\Delta E$ is negative, and accepted with
probability $\exp(-\Delta E/T)$ if $\Delta E$ is positive.

\subsection*{Curvature model}
\begin{figure*}
\includegraphics[width=2.0\columnwidth]{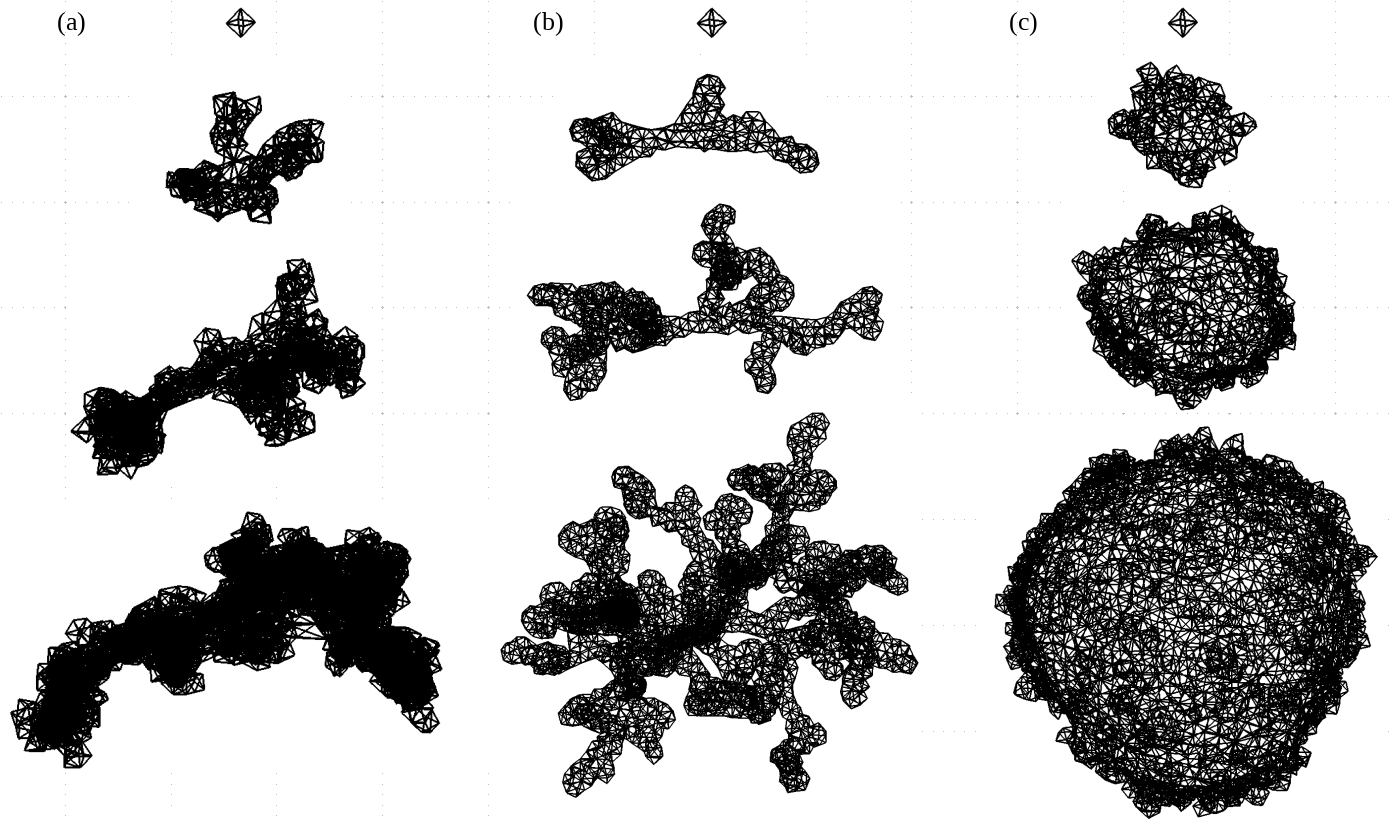}
\caption{\label{hot_and_cold}
\textbf{Growing graphs at hgh and low temperatures; the third column 
shows the main result of this Article: a discrete model of Euclidean space.}
A small octahedral triangulation, with $N=6$ can be grown and 
equilibrated into larger graphs 
with $N=2^{8}$, $2^{10}$ and $2^{12}$ nodes at 
(a) high temperature, 
(b) $T=0.5$ in the curvature model, or 
(c) low temperature in the walker model. The illustrative embedding 
into space shown here is irrelevant to our results; we are only interested 
in the graph.
}
\end{figure*}

The most obvious choice of energy to reduce curvature fluctuations
at low temperature is $E_{\rm curv}=\sum_{i}\kappa_i^2$, where
the sum is over all nodes $i$.
As shown in Figure \ref{hot_and_cold} and also considered in \cite{Aste},
this does indeed drive the local curvature to zero almost everywhere
at low temperature,
but it does so by creating a branched polymer phase consisting
of thin tubes with curvature trapped at their ends and junctions
(Figure \ref{hot_and_cold}b).
The result of this `curvature model' is far from flat. 
We attribute this to
the energy functional failing to sufficiently penalize small 
curvatures spread over large areas.

\subsection*{Walker model}
To address the deficiency of the curvature model, we
introduce a second statistical process by putting walkers on the graph.
Walker models have previously been used to create 
both local-cluster structure \cite{Ikeda,Vazquez,Toivonen},
as well as scale-free \cite{Krapivsky} graphs
from local rules \cite{Saramaki,Caravelli}; but here we are interested 
in Euclidean behavior. At each time step, we
add $\kappa$ walkers of type $+1$ to every node with curvature $\kappa>0$, and
$|\kappa|$ walkers of type $-1$ to every node with $\kappa<0$. 
Additionally, 12 walkers of type $-1$ are added to random nodes 
to maintain the mean walker number [this requirement can be seen 
from eq.\ (\ref{meanZ})].
The walkers then diffuse by moving to a random neighboring node.
Whenever a $+1$ and a $-1$ walker occupy the same node, both 
walkers annihilate. 
Walker moves alternate with push-pop moves. To define the dynamics,
we replace $E_{\rm curv}$ with a new energy $E_{\rm walk}$ for the graph under 
push-pop moves:
\begin{equation}
E_{\rm walk} = -\sum_{i}w_i|w_i|,
\end{equation}
where $w_i$ is the net number of walkers on node $i$.
At low temperatures, this energy
tends to shrink regions of positive curvature and grow regions of
negative curvature.
We call this new evolution scheme, which biases the graph 
towards flatness on long length scales, the `walker model'. 

The walker model generates a triangulation which, at low temperature 
and long lengths, appears
qualitatively to have minimal curvature (Figure \ref{hot_and_cold}c).
To establish that these graphs satisfy Euclidean geometry at long length 
scales, we subject them to three tests: a Hausdorff dimension of 2;
geodesic confinement; and the Pythagorean theorem.

\begin{figure}
\includegraphics[width=1.0\columnwidth]{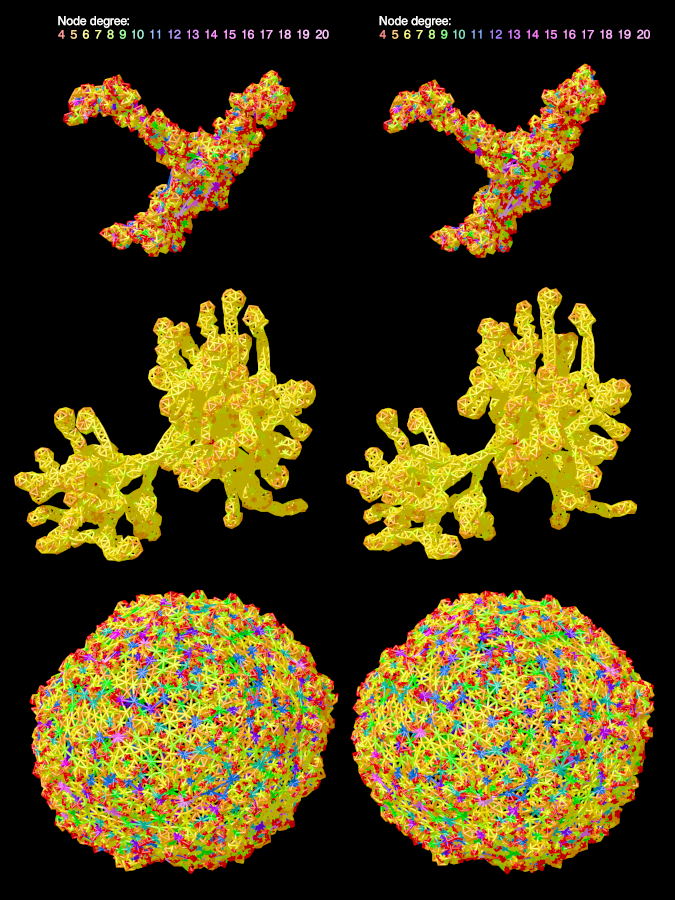}
\caption{\label{stereo}
\textbf{Stereograms of graphs with 6144 nodes.}
Top two images: high temperature graph. Middle two images: curvature model
at $T=0.5$. Bottom two images: walker model at low temperature. The nodes
are coloured according to degree, as shown in the legend at the top of the
Figure. To view as stereograms, the Figure should be held approximately 
30cm away, while looking through 
the page until the two images fuse.
}
\end{figure}

\begin{figure}[t!]
\begin{center}
\includegraphics[width=1.0\columnwidth]{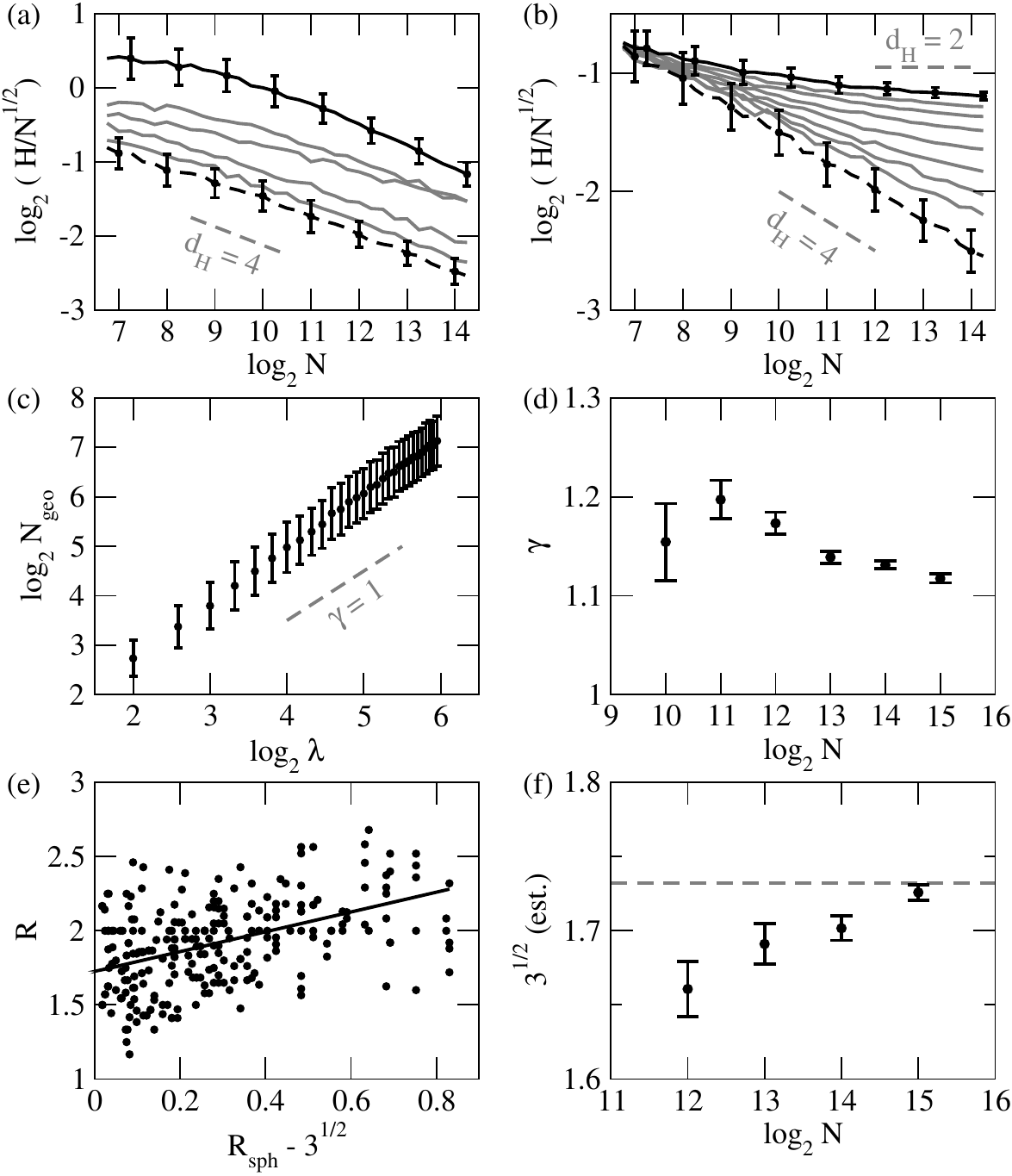}
\end{center}
\caption{\label{all_tests}
\textbf{Statistical tests for Euclidean behavior of our graphs.}
Top row: 
The mean node eccentricity $H$ and standard deviation for example points,
divided by $N^{1/2}$, where $N$ is the number of nodes. (a) The curvature model 
with $T=0.5$ (black), $2^{0}$, $2^{2}$, $2^{4}$, $2^{6}$ (gray) and $10^{5}$
(dashed) (b) The walker model, with $T=2^{-3}$ (black), 
$2^{2}$, $2^{3}$, $2^{4}$ \ldots $2^{8}$
(gray) and $10^{5}$ (dashed).
Middle row:
(c) The number of nodes $N_{\rm geo}$ in geodesic bundles of different lengths
$\lambda$ on a low-temperature walker model graph with $N=2^{15}$ nodes.
(d) Fitted values for $\gamma$, where $N_{\rm geo}\propto \lambda^{\gamma}$ for
graphs of different $N$.
Bottom row:
$R$ is the ratio of the perpendicular length to the edge side of an 
equilateral triangle drawn on a low-temperature walker model with $N=2^{15}$
nodes. $R_{\rm sph}$ is the exact equivalent on a smooth sphere 
[eq. (\ref{great_circle})].
(e) $R$ plotted against $R_{\rm sph}-\sqrt{3}$
(we show a random sample of 250 from the full set of 6078 points).
The line is a linear regression and we
extract the intercept as a graph-theoretic estimate of $\sqrt{3}$.
(f) Estimates of $\sqrt{3}$ by this method for graphs of different size $N$.
The dashed gray horizontal line is the exact value.
}
\end{figure}

\section{Testing our graphs \label{testing}}
Euclidean geometry is defined through five axioms. These are neither as
logically primitive as they first appear, nor do they
readily translate into conditions for discrete models of space. 
We therefore propose three conditions for any discrete model, 
including ours, purporting to capture Euclid's geometry at large 
lengths. The first, Hausdorff dimension, sits outside the original 
axioms, since they concerned the plane. The second condition is the 
appearance of straight lines in the large length limit, which we 
call geodesic confinement. The third is the Euclidean metric itself, 
commonly known as Pythagoras' theorem, which is a synthesis of all
the axioms.

\subsection*{Hausdorff dimension}
If the number of nodes in a 
ball of radius $r$ scales as $N\propto r^{d_H}$, then $d_H$ is
the Hausdorff dimension of the graph. 
As we noted earlier, planarity is not indicative of dimension: random 
triangulations of the 2-sphere have $d_H = 4$ as they converge, in the
large node limit, to `Brownian maps' \cite{leGall}.
To calculate the dimension of our graphs, 
we define the half-circumference $H$ of a graph as the average over all nodes
of the node eccentricity, where the eccentricity of a node is the greatest
geodesic distance between it and any other node in the graph. If nodes
are a measure of area, then we would expect a graph which approximates
a smooth spherical surface with $d_H = 2$ to satisfy the 
scaling $H\propto N^{1/2}$.
This is not the case for the curvature model (Figure \ref{all_tests}a),
but is true for the walker model in the
low temperature limit for a large number of nodes (Figure \ref{all_tests}b).
The upwards curvature 
of the solid gray lines in Figure \ref{all_tests}b shows
evidence that this $d_H = 2$ phase survives to temperatures above zero.

\subsection*{Geodesic confinement}
In a doubly-periodic graph, the total number of nodes $N_{\rm geo}$ in the
geodesic bundle between two nodes a distance $\lambda$ apart
scales as $N_{\rm geo}\propto \lambda^2$.
From Figure \ref{all_tests}cd, we see that the scaling of $N_{\rm geo}$ with
$\lambda$ also approximates a power law for the low-temperature walker model, 
but with a different exponent:
\begin{equation}
N_{\rm geo} \propto \lambda^{\gamma}\ \ {\rm with}\ \ \gamma\approx 1.1.
\end{equation}
An exponent $\gamma<2$ implies qualitatively different behavior to the 
doubly-periodic lattice case, and
in the limit $N\rightarrow\infty$, it is consistent with the narrow 
geodesics (`straight lines') familiar from Euclidean geometry.
We call the collapse of the broad, $N_{\rm geo}\propto \lambda^2$ 
geodesic bundles `geodesic confinement' (Figure \ref{geodesic_pics}b), 
by analogy to the flux tubes and color confinement
seen in strong-force interactions \cite{ColorConfinement}.

\subsection*{Pythagoras' theorem}
Finally, we consider the validity of Pythagoras' theorem on graphs 
generated by the walker model. 
Although this can be proved in general
for Euclidean geometry, on graphs we test it empirically by calculating
the length of the perpendicular of an equilateral triangle. If Pythagoras'
theorem holds, this will be $\sqrt{3}$ times half the side length. 

Because we are generating approximations to a spherical surface, rather than
a plane, we want to make use of as much of the graph as possible, rather than
a small patch on which statistics will be poor. We therefore perform the 
analogous calculation using spherical, rather than plane trigonometry.
If we draw an equilateral spherical triangle on a smooth 2-sphere, 
with side-length $\Lambda$ times the half-circumference, 
the ratio of the length of the perpendicular of the triangle to
half its side length will be
\begin{equation}
R_{\rm sph}(\Lambda)\equiv \frac{2}{\pi\Lambda}\arccos\left[
\frac{\cos(\pi\Lambda)}{\cos(\pi\Lambda/2)}\right]=
\sqrt{3}+O(\Lambda^2).
\label{great_circle}
\end{equation}
The same ratio $R$ can be calculated for a graph formed by the
low-temperature walker model (Figures \ref{all_tests}ef,
\ref{equilateral}), 
and although the fluctuations
are significant, they appear to be unbiased, so that
performing linear regression of $R$ against $R_{\rm sph}$ gives an 
estimate for $\sqrt{3}$ which is only
one standard deviation from the traditional value:
\begin{equation}
\sqrt{3}_{\rm est} = 1.726 \pm 0.005.
\end{equation}
We believe this is a non-trivial result, unlikely to emerge accidentally,
and so we take it as strong evidence that Pythagoras' theorem is
satisfied in general for the low-temperature walker model. Since the
straightedge and compass operations of drawing circles of any radius,
and drawing and measuring (but not extending) lines are simple operations
on graphs, many other constructions of classical geometry may readily be tested.

\section{Methods}
Our proof that all planar lattice graphs satisfy the taxicab metric
is in two parts, which we call geodesic composition and geodesic rearrangement.

\subsection*{Proof of geodesic composition}
Consider a doubly-periodic planar graph made up of identical unit cells, 
each of which comprises $\omega$ distinct nodes.
Equivalent nodes in different unit cells are said to be of the same type.
Let ${\cal G}_{pp}({\rm\bf v})$ denote a particular geodesic between 
two $p$-type nodes separated by $\mathbf{v} = (m,n)$ unit cells.

\begin{figure}
\includegraphics[width=1\columnwidth]{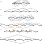}
\caption{
\textbf{All doubly-periodic planar graphs have a taxicab metric on long 
length scales.}
(abcd)
A grid of unit cells forms a doubly-periodic planar graph;
nodes within the unit cells not shown.
For some node type $p$, if $p_0 p_1$ is a shortest path between 
nodes separated by $\mathbf{v} = (m,n)$ unit cells,
then $p_0 p_1 \ldots p_k$ is the shortest path between 
nodes separated by $k \mathbf{v}$ unit cells.
(ef)
For $m$ and $n$ relatively prime, 
the geodesic ${\cal G}_{pp}(k{\rm\bf v})$ is the concatenation of 
$k-1$ copies of both ${\cal G}_{qq}({\rm\bf u})$ and
${\cal G}_{qq}({\rm\bf u'})$, with tails at either end.
See the text for details.  
}
\label{taxiproof}
\end{figure}

We first prove that for any displacement $\mathbf{v}$, for at 
least one node type $p$, the concatenation ${\cal G}_{pp}(k{\rm\bf v})$  
of $k$ copies of ${\cal G}_{pp}({\rm\bf v})$ is also a geodesic
(Figure \ref{taxiproof}a--d).
Let $p$ be the node type which minimizes ${\cal G}_{pp}({\rm\bf v})$. 
Call this the optimal node assumption.
Let $p_0 p_1$ of length $\vert p_0 p_1 \vert = \lambda$ be a geodesic 
between $p_0$ and $p_1$, which are both of type $p$, but
displaced ${\rm\bf v}$ units cells from one another (Figure \ref{taxiproof}a).
Call this the $\mathbf{v}$-geodesic assumption. Let $p_0 p_1 p_2$ be
two copies of $p_0 p_1$ placed end-to-end.

Now suppose there is a path $p_0 a b p_2$ with 
length $\vert p_0 a b p_2 \vert < \vert p_0 p_1 p_2 \vert = 2 \lambda$ 
(Figure \ref{taxiproof}b);
because the graph is planar, nodes $a$ and $b$ exist.
Then $\vert a b \vert  < \lambda$ or 
$\vert p_0 a \vert + \vert b p_2 \vert < \lambda$.
If the former, then we contradict the optimal node assumption.
If the latter,  we contradict the $\mathbf{v}$-geodesic assumption.
Therefore $p_0 p_1 p_2$ is a geodesic between $p_0$ and $p_2$. That is to say,
${\cal G}_{pp}(2{\rm\bf v})$, which is the concatenation of 2 copies 
of ${\cal G}_{pp}({\rm\bf v})$, is a geodesic.
Call this the $2 \mathbf{v}$-geodesic property.

\begin{figure}
\includegraphics[width=1.0\columnwidth]{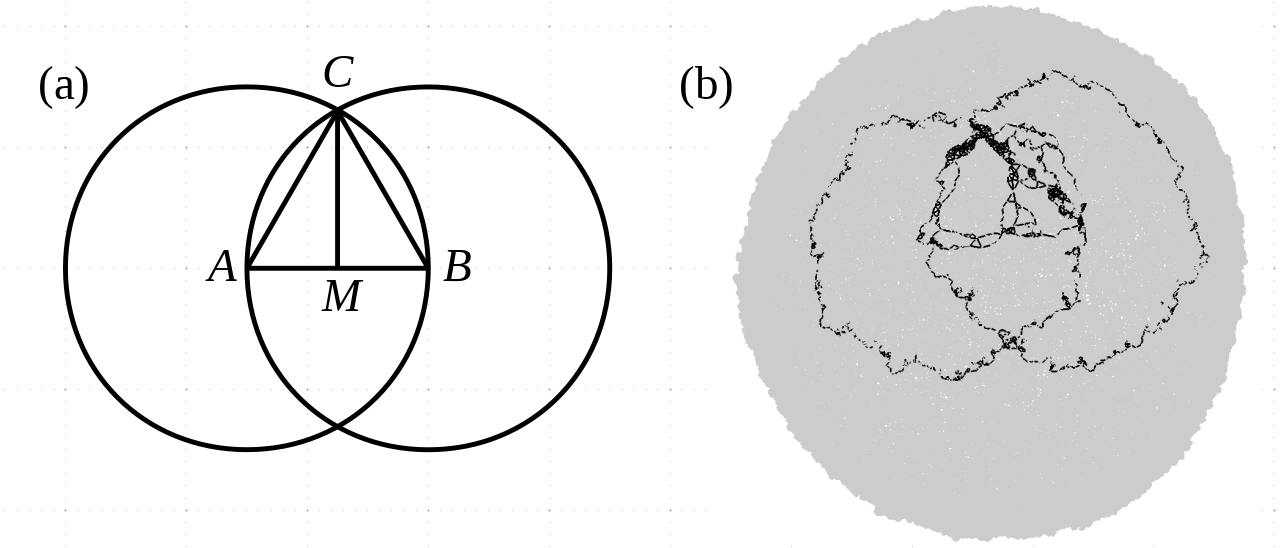}
\caption{\label{equilateral}
\textbf{Equilateral triangles on the plane and on a graph.}
(a) An equilateral triangle drawn on the Euclidean plane with
straightedge and compass, where
$M$ is half-way between $A$ and $B$, and $MC/AM=\sqrt{3}$.
(b) The same construction using geodesics on a low-temperature `walker model'
graph (which approximates a smooth sphere)
with $N=2^{16}$ nodes and triangle side length of 32.
}
\end{figure}

We now show that the $(k-1) \mathbf{v}$-geodesic property
implies the $k \mathbf{v}$-geodesic property
(Figure \ref{taxiproof}c for $k=3$).
Suppose there is a path $p_0 a b p_k$ with 
length $\vert p_0 a b p_k \vert < \vert p_0 p_1 \ldots p_k \vert =  k \lambda$.
Then $\vert a b \vert  < \lambda$ or $\vert p_0 a \vert + \vert b \, p_k \vert < (k-1) \lambda$
(Figure \ref{taxiproof}d for $k=3$).
If the former, then we contradict the optimal node assumption.
If the latter,  then we contradict the $(k-1)\mathbf{v}$-geodesic property.
Therefore $p_0 p_1 \ldots p_k$ is a geodesic between $p_0$ and $p_k$.
This completes the first part of the proof.

\subsection*{Proof of geodesic rearrangement}
We next prove that for most displacements $\mathbf{v}$, for at least one 
node type $p$, the geodesic ${\cal G}_{pp}(k{\rm\bf v})$ 
consists of three parts: a tail at each end, which joins the
nodes $p_0$ and $p_k$ to copies of some other type of node $q$, 
and between the tails, $k-1$ alternating copies of
${\cal G}_{qq}({\rm\bf u})$ and ${\cal G}_{qq}({\rm\bf u'})$ for
some displacement vectors ${\rm\bf u}$ and ${\rm\bf u}'$
(Figure \ref{taxiproof}ef).
We now only consider displacement vectors ${\rm\bf v}=(m,n)$ such that
$m$ and $n$ are relatively prime (which occurs \cite{Hardy} for random $m$ and
$n$ with probability $6/\pi^2 \simeq 0.61$) and large enough 
so that $\lambda > 2 \omega$,
where $\omega$ is the number of distinct nodes in the unit cell.
By Dedekind's pigeonhole principle \cite{Pigeon}, 
since $\lambda/\omega > 2$, ${\cal G}_{pp}({\rm\bf v})$ must pass 
through at least two nodes of some other type $q$ different from type $p$
(Figure \ref{taxiproof}e).
Therefore we can define a sub-geodesic
${\cal G}_{qq}({\rm\bf u})$ within ${\cal G}_{pp}({\rm\bf v})$, and
a second geodesic ${\cal G}_{qq}({\rm\bf u}')$ between the node $q$ in
adjacent copies of ${\cal G}_{pp}({\rm\bf v})$ (Figure \ref{taxiproof}f).

Because $m$ and $n$ are relatively prime, ${\rm\bf u}$ 
and ${\rm\bf u}'$ cannot be parallel.
To see why, let the displacement ${\rm\bf u}$ be $(i,j)$ and the 
displacement ${\rm\bf u}'$ be $(i',j')$ and assume $i' \ge i$.
Since ${\rm\bf u} \parallel {\rm\bf u}'$ 
implies $i/j = i'/j'$, $(m,n) = (i+i',j+j') = (1+i'/i) (i,j)$, 
where $i'/i$ is an integer, contradicting $(m,n)$ being relatively prime.

The $k-1$ alternating geodesics can be rearranged in any order, 
forming a set of staircases between the end $q$ nodes (Figure \ref{taxiproof}f).
Therefore the geodesic bundle occupies an area of $m \, n (k-1)^2$ unit cells.
This completes the proof.

\subsection*{Computer code}
See Supplemental Material at [URL will be inserted by publisher] for 
the computer code (in the C progamming language) used to generate
the data in Figure \ref{all_tests}, as well as the pctures in
Figures \ref{hot_and_cold}, \ref{stereo} and \ref{equilateral}.
There is also a video showing the different network models' dynamics.

\section{Discussion}

We have shown that discrete space and Euclidean space, thought by 
many to be at odds, are indeed compatible. We avoid Zeno's 
paradox because we do not require our model to be infinitely 
divisible. We avoid Weyl's tiling argument because our model is 
disordered. Weyl's argument is in fact an observation that certain 
non-planar lattices display the taxicab metric, which is unsurprising 
given our proof that all planar lattice graphs do.

Our model shows the emergence of Euclidean space on long length scales,
from a local statistical process that employs only the intrinsic geometry
of a graph. Beyond providing the first discrete model of the Euclidean plane,
we believe our results draw together several disparate fields of physics,
and prompt many intriguing questions that merit further investigation.

\subsection*{No embedding space} 

Smooth surfaces which are discrete at an atomic scale frequently 
arise in nature, such as liquid menisci or crystal surfaces
\cite{Burton}. These atomic systems 
are embedded in a background manifold, consisting of ordinary,
flat, three-dimensional space.
This embedding manifold allows distance on the surface to be defined
in the usual Euclidean manner, and also means that normals to the
surface exist. The system energy can then depend on extrinsic curvature
(the spatial gradient of these normals), as well as intrinsic (Gaussian) 
curvature.  Our graphs, by contrast, do 
not live in a background space. Instead, our measures of distance 
and curvature are only intrinsic, defined in terms of edges (distance)
and node degree (curvature) that are properties of the graph itself. 
No normal vectors to our graph manifolds exist, nor can they be defined.
Our graphs are themselves the space, and their edges are the quantum of
distance.

We note that because our graphs approxmate boundary-less
two-dimensional manifolds, we can approximate the plane only by 
forming a very large 2-sphere (since a large, smooth
2-sphere is locally close to a plane), or by forming a 
topological 2-torus. Our model in its current form is not able to grow any
manifold with a boundary, such as a disc.

\subsection*{Phase transition} 

Phase transitions which create or destroy smoothness are well 
known in physics. A roughening transition \cite{Burton} can turn flat
crystal facets into smooth, curved surfaces, as measured with
the metric of the embedding space. However, this embedding space is needed
to define what smooth curvature means in this case. 

More strikingly, the crumpling transition of membranes \cite{Bowick} turns
flat crystalline membranes into crumpled objects, confined in
a small region of space. However, the
irregular, jagged curvature of the crumpled phase is entirely extrinsic:
a function of its embedding in three-dimensional space.
The intrinsic, ordered, taxicab geometry of the membrane itself
is unchanged through the crumpling transition.

In contrast, the phase transition we find at low temperature 
in the walker model changes the {\em intrinsic} metric of the graph
from a crumpled, non-Euclidean `Brownian map'
\cite{leGall} into smooth, Euclidean space.
It is unclear, at present, whether the phase transition occurs at
finite or infinite temperature.
A renormalization group analysis of the model may shed light on this question.

\subsection*{Walker model} 

The phase transition which creates continuum geometry is driven by
a statistical walker process. The motivation for this comes
from the na\"{i}ve curvature model, which minimizes the sum of the squares
of the local discrete curvature $\kappa$, but disappointingly 
gives rise to a `Medusa' phase 
(Figure \ref{hot_and_cold}b). This pathological behavior
is consistent with previous investigations 
of triangulations, which lead to branched polymer phases and other 
exotic geometries rather than smooth, homogenous space \cite{Aste,Gurau}. 
The pathologies are due to concentrations of discrete curvature
in confined regions; in other words, large, local curvature fluctuations. 
Our walker process -- which solves a discrete 
version of Poisson's equation, with the charge being the 
curvature $\kappa$ -- is sensitive to small curvatures on large length scales, 
and so, through the energy functional $E_{\rm walk}$, ultimately acts to 
spread these fluctuations over the whole graph.

\subsection*{A background for simulations}

A practical application of our Euclidean graphs is as a background 
for simulations. Lattices, such as the square grid, are intrinsically 
anisotropic, so special care is often needed when designing simulations
to run on them. The rotational symmetry of our 
graphs therefore make them suitable spaces on which to perform algorithms
such as lattice gas cellular automata \cite{LGCA}.

\subsection*{Higher dimensions} 

We have built a discrete, graph model that behaves like two-dimensional 
Euclidean space at large lengths. Can the same be done for 
higher dimensions? While more computationally intensive,
our walker model should generalize naturally to dimensions greater than two.
In three dimensions, the key step is extending the Steinitz moves in 
Figure \ref{moves} to add and subtract tetrahedra, rather than
triangles, as nodes divide and fuse.
Whether the resulting graph will be Euclidean is, however, unknown.  
Our tests for geodesic confinement and the applicability of 
Pythagoras' theorem are benchmarks for this and any other 
discrete models attempting to capture Euclidean geometry at large lengths.

We conjecture that the absence of geodesic confinement 
carries over to higher dimensional lattices, as it clearly does 
for the three-dimensional regular cubic grid. 
Unfortunately, the proof does not readily 
follow from our theorem in two dimensions, which relies on planarity, since 
all three-dimensional lattices are non-planar. Figure \ref{geodesic_pics}c 
gives an indication of the subtlety. It shows a non-planar, two-dimensional
lattice not satisfying
geodesic composition, a key step in our proof (see Methods).

Closely related results have been proved for more general cases.
It is known that for asymptotically large radii, balls around
a node in a lattice graph of any dimension can never be ellipsoids,
but are rational polytopes \cite{Fritz}. Isotropic behavior of
geodesics is therefore impossible, and if these polytopes were exact,
rather than having finite length whiskers in certain directions (the
eventuality of Figure \ref{geodesic_pics}c) broad geodesic bundles would
follow from the partial incidence of the faces of polytopes of radius
$r$ around two points separated by $2r$.

\subsection*{The Minkowski metric} 

We have shown how to grow graphs with a Euclidean metric, 
that is, that satisfy Pythagoras' theorem, $d^2 = x^2 + y^2$,
where $d$ is distance and $x$ and $y$ orthogonal directions.
What about other metrics? The most important is
the Minkowski metric of special relativity, the two-dimensional analog 
of which is  $d^2 = t^2 - x^2$, where $t$ is a time direction.
How to represent this as a 
graph is an open question, because nodes must be intricately
connected at large coordinate displacements. Taking an approach similar
to causal set theory \cite{Bombelli,Benincasa}, but with neighbors
separated by unit proper time,
would suggest that the degree of each node diverges with the
logarithm of the volume of space-time (or worse, as a power, for 
higher dimensions).
Furthermore, unlike Euclidean space, where the square grid 
graph at least models a
4-fold rotational symmetry, it is not possible 
to construct a lattice graph which is symmetric under 
even a discrete version of the Lorentz transformation.
Thus, it remains to be seen whether some variant of 
the walker process can be defined to probe and
engender the fabric of space-time.

\begin{acknowledgements}
The authors thank Prof.\ Yang-Hui He and Dr Ilia Teimouri for useful 
discussions. 
\end{acknowledgements}


\end{document}